\documentclass[granma]{svjour}
\usepackage[latin1]{inputenc}

\usepackage{graphicx}% Include figure files
\usepackage{dcolumn}% Align table columns on decimal point
\usepackage{bm}% bold math

\usepackage{color}
\newcommand{\rev}[1]{#1}

\begin{document}

\title{A micromechanical model of collapsing %% soil/living
  quicksand}

\author{Dirk Kadau\inst{1}\mail{dkadau@ethz.ch}, 
Jos\'e S. Andrade Jr.\inst{2} and Hans J. Herrmann\inst{1,2}}

% \thanks{We thank Sarah Springman for lending us her equipment and CNPq, CAPES,
% FUNCAP, FINEP, the Volkswagenstiftung and the Max Planck Prize for
% financial support.
%  }

\institute{\inst{1}Institute for Building Materials, ETH H\"onggerberg,
8093 Z\"urich, Switzerland\\ \inst{2}Departamento de F\'{\i}sica,
Universidade Federal do Cear\'a, 60451-970 Fortaleza, Cear\'a,
Brazil}

\date{\today}

\maketitle

\begin{abstract}

The discrete element method constitutes a general class of modeling techniques
to simulate the microscopic behavior (i.e.\ at the particle scale) of
granular/soil materials. We present a %% variant of the
contact dynamics method,
accounting for the cohesive nature of fine powders and soils. 
A modification of the model adjusted to capture the essential physical
processes underlying the dynamics of generation and collapse of loose systems is
able to simulate "quicksand" behavior of a collapsing soil
material, in particular of a specific type, which we call ``living
quicksand''.
We investigate the penetration behavior of an object for varying density of
the material. We also investigate the dynamics of the penetration process, by
measuring the relation between the driving force and the resulting velocity of
the intruder, leading to a ``power law'' behavior with exponent $1/2$, i.e.\ a
quadratic velocity dependence of the drag force on the intruder.

\end{abstract}

\section{Introduction}

%% \begin{itemize}
%% \item more generally introduction, also refer to complex fluids as velocity
%%   dependence ??
%% \end{itemize}

Despite the ubiquitous appearance of quicksand in adventure books and
movies, its origin and physico-chemical behavior still represent
controversial scientific issues in the fields of soil and fluid
mechanics \cite{Freundlich35,Matthes53,Krus96,Bahlmann02,Yamasaki03}. It has been argued repeatedly
\cite{Smith46} that, because the density of sludge is typically larger
than that of water, a person cannot fully submerge, and therefore
cannot be really ``swallowed'' by any {quicksand}.

The fluidization of a soil due to an increase in ground water
pressure, which in fact is often responsible for catastrophic failures
at construction sites, is called by engineers the {``quick condition''}
\cite{Craig97,Shamy05}, and has been studied extensively up to now
\cite{Vardoulakis1989,Vardoulakis2004a,Vardoulakis2004b}. % 1989 also model
Another source of fluidization can be
vibrations either from an engine \cite{Huerta2005} or through an
earthquake~\cite{Krus96}. While this ``liquefaction'' \rev{ or ``cyclic
  mobility'' phenomenon \cite{Castro75,Ishihara93,Pastor90}  }  can
essentially happen with any soil \cite{Lambe69}, it is known that
samples taken from natural quicksand usually show quite specific, but
anomalous rheology depending on the peculiarities of the material
composition and structure \cite{Khaldoun05}. {Also in dry
  quicksand \cite{Lohse04,Royer2005} a fragile/metastable 
  structure leads to interesting material behavior  \rev{like collapse and jet
    formation after impact of an intruder}, although the microscopic
  material properties are obviously quite different. Nevertheless, wet and dry
  quicksand are more similar than one would think.}   Based on results of real and {\it in situ} quicksand measurements we develop and numerically solve a
modified version of the previously presented simulation of a simple
physical model for this {quicksand}/collapsing soil.  Here, we focus on the
penetration behavior of {an intruder}.

\section{Experiments and Model}

% \subsection{Experiments}

 Our physical model is inspired by in situ measurements performed with a
specific type of natural quicksand at the shore of drying
lagoons located in a
natural reserve called Len\c{c}ois Maranhenses in the
North-East of Brazil \cite{kadau2009a,kadau2009b,kadau2009c,kadau2010a}.
Cyanobacteria form an impermeable crust, giving the impression of a 
stable ground. After breaking the crust a person rapidly sinks to the bottom
of the field. We measured the shear strength of the material before and after
perturbation and found a drastic difference. 
Our measurements indicate that
the quicksand is essentially a collapsing suspension with depth independent shear strength. After the
collapse, it becomes a soil dominated by the Mohr-Coulomb friction
criterion for its shear strength. The material undergoes a cross-over
  from a yield stress material, i.e.\ a more fluid-like behavior to a Coulomb
  material, i.e.\ more solid-like behavior after the collapse. 
We would like to point out that the collapse of the metastable structure
is irreversible, as opposed to
quicksand described in Ref.~\cite{Khaldoun05}.
In summary, the ``living quicksand'' studied here can be described as a
suspension of a tenuous granular network of cohesive particles. If
perturbed, this unusual suspension can drastically collapse, promoting
a rapid segregation from water, to irreversibly bury an intruding
object.

%\subsection{Simulation}

To model the complex behavior found in the experiments we use a variant of
contact dynamics, originally developed to model compact and dry systems
with lasting contacts \cite{Moreau94,Jean92}.  %\cite{Moreau94,Jean92,unger2003,brendel2004}.
The absence of cohesion between particles can only be justified in dry
systems on scales where the cohesive force is weak compared to the
gravitational force on the particle, i.e. for dry sand and coarser
materials, which can lead to densities close to that of random dense
packings. However, an attractive force, \rev{ e.g.\ due to capillary bridges
  or van 
der Waals forces,}  plays  an important role in the stabilization of large voids
\cite{Kadau03}, leading to highly porous systems as e.g. in fine
cohesive powders, in particular when going to very small grain
diameters. %% In the nanometer range of particle sizes, the cohesive
%% force becomes the dominant force, so that particles stick together
%% upon first contact.
Also for contact dynamics a few simple models
for cohesive particles have been established
\cite{taboada2006,richefeu2007,Kadau03}. Here we consider the bonding between two particles  in terms of a cohesion model with a constant attractive
force $F_c$ acting within a finite range $d_c$, so that for the
opening of a contact a finite energy barrier $F_cd_c$ must be
overcome. In addition, we implement rolling friction between two
particles in contact, so that large pores can be stable
\cite{Kadau03,kadau_brendel_2003,bartels_unger_2005,brendel_kadau_2003,morgeneyer_roeck_2006}.

In the case of collapsing ``living quicksand'' our cohesion model has to be modified. One also has to take into
account the time necessary for bonds to appear, i.e.\ during
relatively fast processes new bonds will not be formed, whereas during
longer times bonds are allowed to form at a particle
contact. Finally, gravity also cannot be neglected in the model since
the particle diameter is usually well above the micron-size. For
simplicity, however, the surrounding pore water is not explicitly
considered but only taken into account as a buoyant medium, reducing
the effective gravity acting onto the grains.
\rev{ Disregarding the interstitial fluid motion keeps the model as simple as
  possible and nevertheless able to reproduce the main experimental
  observations \cite{kadau2009b}. The details of the collapse, however, may be
influenced by the flow field of the surrounding fluid
\cite{Royer2005,Caballero2007}. Testing the influence of the fluid by
introducing a viscous drag onto the grains considering water and the typical grain sizes from the experiments showed no significant difference
\cite{kadau2009c,kadau2010a}. } 
 As previously discussed, we justify the cohesive
bonds in this case as being mediated by the bacteria living in the
suspension.

Summarizing, we use the following contact laws for the simulations
  presented in this paper. Perfect volume exclusion (Signorini condition) is
  assumed in normal direction, where a cohesive force is added as described
  above. In tangential direction Coulomb friction is applied (Coulomb
  law). Additionally, for the contact torques rolling friction is applied as a
  threshold law similar to the Coulomb law. A detailed description of this
  model can be found in Ref.\ \cite{Kadau03}. For the simulation of the collapsing
  living quicksand the formation and breakage of cohesive bonds has to be
  considered additionally. 
 Our physical model is
validated with the real data obtained from the situ measurements, specifically
by comparing the shear strength behavior \rev{showing a drastic change when soil
was perturbed}  and the penetration
behavior \rev{ described in more detail in Refs.\ }  \cite{kadau2009a,kadau2009b,kadau2009c}. 

\section{Results}

The penetration behavior of our collapsing ``living quicksand'' has
been studied previously for one specific density
\cite{kadau2009a,kadau2009b,kadau2009c,kadau2010a} and is here only briefly
summarized. The penetration
causes the partial destruction of the porous network and the
subsequent compaction of the disassembled material. We observe the
creation of a channel which finally collapses
over the descending intruder. At the end of the penetration process
the intruder is finally buried under the
loose debris of small particles. Furthermore, our simulations indicate that in
the worst condition, \rev{when the cohesive force is completely restored}, one
could need a force up to three times one's weight to get out of such
morass \cite{kadau2009c}. 
In this paper we will focus on the penetration process for varying densities
while keeping the cohesive force constant.

\begin{figure}[htb]
\vspace{-5ex}
\begin{center}a) \includegraphics[%
  width=0.37\columnwidth]{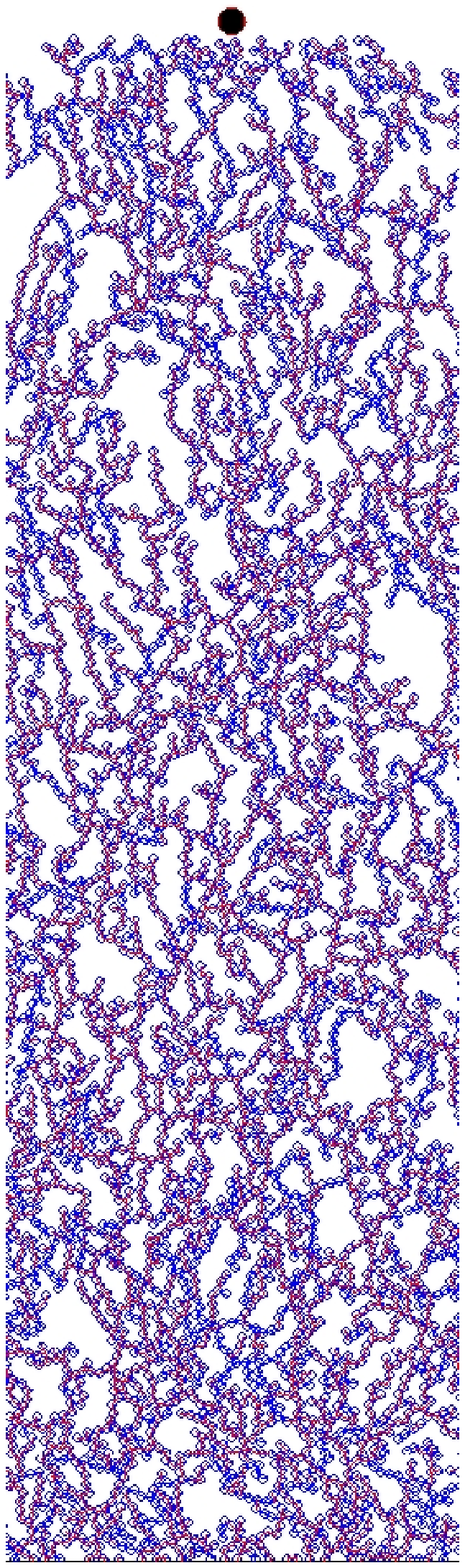} b) \includegraphics[%
  width=0.37\columnwidth]{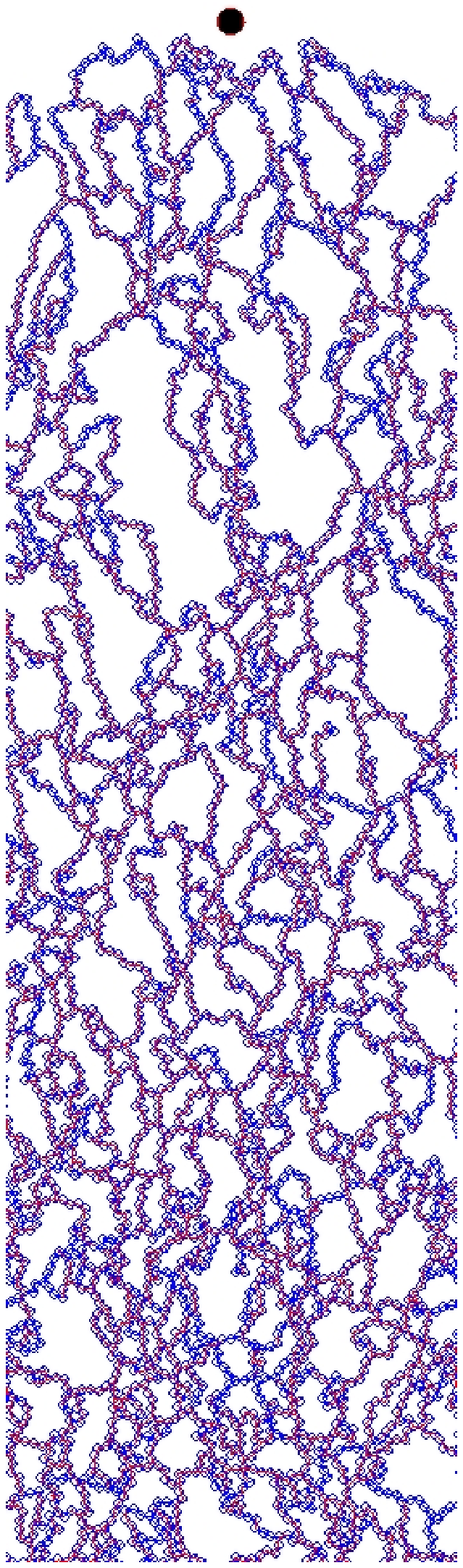}\end{center}

\caption{\label{fig:snaps_min_max} Maximal density of a configuration after
  settling without change (a) and minimal density after removing all possible
  loose ends (b).}
\end{figure}

%\subsection{Fragile structures with specified density}\label{sec:ini}
%
When creating fragile structures by ballistic deposition and settling due to
gravity the density is determined by the strength of the cohesive force. Here
we will show a way of generating fragile structures with determined density
(within a given range). The particles are deposited ballistically, and
settle due to gravity.
 After the settlement of all particles, the
cohesive forces between them are tuned to the point in which a barely
stable structure of grains is assured. This results in a tenuous network of
grains, like in a house of cards (see fig.\ \ref{fig:snaps_min_max}a), 
%
 %% This defines
 giving the maximal possible density for this specific configuration. In
 the case shown the volume fraction is $0.441$. For different
 configurations (different seed for the random number generator), used for later
 averaging, this maximal volume fraction has values ranging from $0.428$ to
 $0.451$. %% The configurations differ only in the seed for the random number
%%  generator used for the ballistic deposition.

For averaging we need configurations with the same initial density. How can we
create such structures? Due to the ballistic deposition there are many
loose ends in the structure, which do not carry any load. When elimination
those loose ends successively one can reduce the density to a given value.
This eliminating procedure will be briefly discussed in more detail. Particles
with only one contact are not contributing to the force network carrying the
weight of the material. Thus, these particles can be eliminated without
leading to a collapse of the structure. We check for each particle (in 
random order) if they have only one contact in which case we eliminate the particle. After
we went through all the particles we start again checking all particles, as
there will be new particles with coordination number one due to the elimination of a
former neighboring particle. This can be done until one reaches a state where
no loose ends are present any more. This defines the minimal possible density
for a specific configuration as shown in fig.\ \ref{fig:snaps_min_max}b,
with volume fraction $0.344$ for this specific case.  For different
configurations this minimal volume fraction ranges from $0.327$
to $0.35$. This process can be interrupted when a desired density is
reached.     

%\subsection{Penetration depth}

\begin{figure}[htb]
\begin{center}\includegraphics[%
  width=0.8\columnwidth]{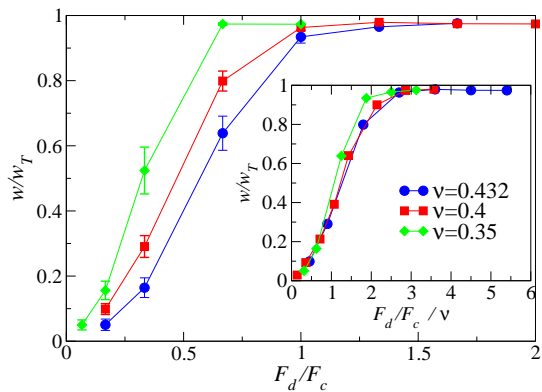}\end{center}
\caption{\label{fig:depth}\label{fig:scaling} Penetration depth \rev{here measured by the weight
  $w$ of 
  all grains above the intruder, i.e. larger vertical position, normalized by
  the total weight $w_T$) depending on the force $F_d$
  acting on   the intruder (normalized by cohesive force $F_c$)}  for
different initial densities of the material. (Connecting 
  lines are shown here for different curves to be better
  distinguishable.) \rev{ Inset: Scaling the force axis  by the volume fraction of the initial configuration leads to a relatively
  good collapse for low values of $F_d$. For higher forces the curves show
  small deviations due to inertial effects.} }
\end{figure}
%
%% \begin{figure}[htb]
%% \begin{center}\includegraphics[
%%     width=0.8\columnwidth]{bilder/weight_vs_force_dens_scaling_lin.eps} 
%% \end{center}   
%% \caption{\label{fig:scaling} Scaling the force axis of fig.\ \ref{fig:depth}
%%   by the volume fraction of the initial configuration leads to a relatively
%%   good collapse for low values of $F_d$. For higher forces the curves show
%%   small deviations due to inertial effects. }
%% \end{figure}
% 
 Fig.\  \ref {fig:depth} shows the penetration depth depending on the applied
 force for different densities of the initial configuration.
%%  For volume
%%  fraction $0.432$ the results obtained for averaging different configurations
%%  (see sec.\ \ref{sec:ini}) 
%%  is compared to the result where the intruder position is varied using the
%%  same configuration. The two curves agree within the error bars.\\
%% {\footnotesize
%% (Here, the
%%  probable error, i.e.\ mean fluctuations divided by $\sqrt{n-1}$ as one
%%  expects a higher accuracy with increasing sample number $n$, see e.g.\
%%  Gould Tobochnik, chapter 11.4.)} %% also lecture M. Troyer
Reducing the density the threshold, needed for pushing in the intruder,
reduces, as 
one expects.  Scaling the force with the density, i.e.\ dividing the
$x$-axis by the volume fraction leads to a data collapse for small forces
only (Fig.\ \ref{fig:scaling}, inset). In this regime where the intruder is
relatively slow, the density fully determines how deep the intruder can be
pushed in.  For higher forces the collapse gets worse due
to inertial effects. In this region a better collapse could be achieved when
scaling with the square of the density. The influence of inertia can be
understood as follows:  Structures with larger densities strongly hinder the
motion of the
intruder. Thus, inertial effects are less effective for these
structures, leading to lower penetration depth within these structures.

%\subsection{Velocity of intruder} \label{sec:velocity}

%
%\subsection{Time dependence of velocity} \label{sec:velocity_time}

%% \begin{figure}[htb]
%% \begin{center}a)  \includegraphics[%
%%  width=0.44\columnwidth]{bilder/ypos_fit_0.432_diffpos_down7e4.eps}
%%  b) \includegraphics[%
%%   width=0.44\columnwidth]{bilder/ypos_fit_0.432_diffpos_down2e4.eps}
%% \end{center}

%% \caption{\label{fig:pos_time} Vertical position depending on time for forces
%%   above the  threshold (here: $F_d/F_c=2.17$) (a) and at threshold
%%   ($F_d/F_c=0.67$)  (b). }
%% \end{figure}

%% \begin{figure}[htb]
%% \begin{center}a)  \includegraphics[%
%%  width=0.44\columnwidth]{bilder/vy_0.432_diffpos_down7e4.eps}
%%  b) \includegraphics[%
%%   width=0.44\columnwidth]{bilder/vy_0.432_diffpos_down2e4.eps}
%% \end{center}

%% \caption{\label{fig:vel_time} Vertical velocity (cf.\  fig.\ \ref{fig:pos_time}) depending on time for forces
%%   above the  threshold (here: $F_d/F_c=2.17$) (a). and at threshold
%%  ($F_d/F_c=0.67$)  (b).
%%  }
%% \end{figure}
When looking in more detail at the dynamics of the penetration process it can
be seen that, after an initial phase, the velocity
fluctuates around a constant value (not shown here) before finally decelerating
to zero at the final intruder position, i.e.\ at the bottom, or at
intermediate values. %% , for values close to the threshold.
%%  the final intruder position varies for 
%% different runs still showing a period of constant velocity (fig.\
%% \ref{fig:pos_time}b and   \ref{fig:vel_time}b).
%% This constant average velocity $v_{\rm avg}$ can be measured and will be
%% investigated in more detail in sec.\ \ref{sec:vel_avg}. 
%% Note, that for small forces the region of constant velocity cannot be
%% found.
%\subsection{Average velocity} \label{sec:vel_avg}
%
%\subsubsection{One specific density}
%
\begin{figure}[htb]
\begin{center}  \includegraphics[%
  width=0.8\columnwidth]{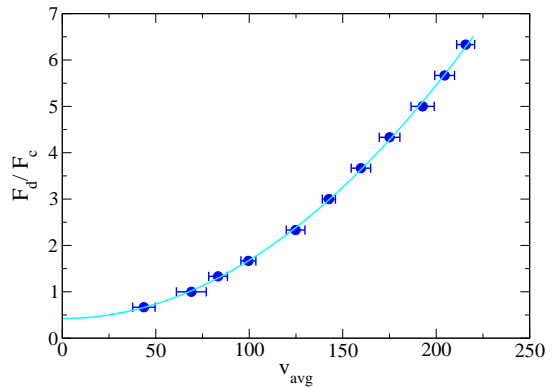}
\end{center}
\caption{\label{fig:force_vel} Relation between applied force and measured
  average velocity when pushing downwards at one specific density (%% volume
%%   fraction 
$\nu=0.432$). A parabolic fit represents the data
  quite well.}
\end{figure}
For different forces applied on the intruder these average velocities can be
measured. The relation between
the applied force and the measured velocity can be fitted by a quadratic
function (fig.\ \ref{fig:force_vel}), where the minimum of the parabola
is very close to $v_{\rm avg}=0$. Alternatively, this minimum can be fixed
at  $v_{\rm avg}=0$ leading to an almost identical fitting curve.  
Summarizing, we found:
\begin{equation}
  \label{eq:force_vel}
  F_d-F_{\rm thr} \propto v^2_{\rm avg}
\end{equation}
In this case the value for $F_{\rm thr}/F_c$ is around $0.42$ which is in
accordance to the value one could estimate from the force dependence of the
relative weight $w/w_T$ above the intruder after penetration. 
%%  Considering the applied force equalizing the drag force on the intruder
%% exerted by the surrounding ``complex fluid'', a square velocity dependence
%% usually implies that inertia effects are important.
\rev{ Let us consider the applied force to equalize the drag force on the
  intruder exerted by the surrounding ``complex fluid''. The results indicate
  a yield-stress fluid behavior as also found in shear strength
  measurements \cite{kadau2009c,kadau2010a}. Future rheological measurements
  could define the specific rheological model.   A
  square   velocity dependence of the drag force }
 usually implies that inertia effects are important. 
Here, the intruder has to
accelerate the grains as they are pushed downwards within the compaction
process. On the other hand, the force is also needed to break cohesive bonds.
The fact that we do not find a viscous like behavior (linear in velocity)
agrees with the non-reversibility of the system.

\begin{figure}[htb]
\begin{center}  \includegraphics[
    width=0.8\columnwidth]{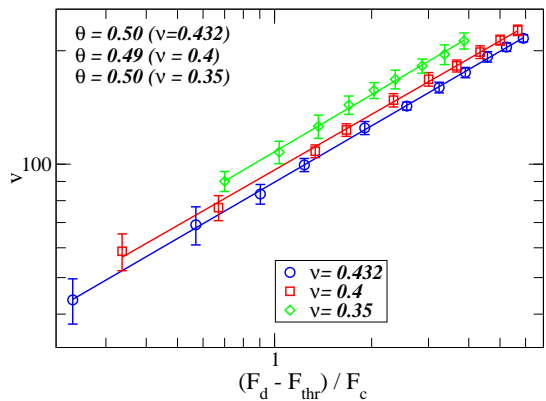} 
\end{center}
\caption{\label{fig:force_vel_log} When plotting the average velocity versus
  the applied driving force $F_d$  minus a threshold force $F_{\rm thr}$,
  a power law fit has an exponent of  around $1/2$ for the three different
  investigated volume fractions.}
\end{figure}

A link can be drawn to the pinning-depinning transition for a force
driven interface where the driving force $F_d$ has to overcome a threshold
$F_{\rm thr}$ leading to a power law: 
\begin{equation}
  \label{eq:pinning}
  v\propto (F_d-F_{\rm thr})^{\theta}
\end{equation}
The former quadratic behavior previously observed would lead to an exponent
$\theta=1/2$. The results obtained for different
densities are shown in fig.\ref{fig:force_vel_log}. 
Note that one has to determine $F_{\rm thr}$ by adjusting to the
optimal power law behavior, then a power law fit estimates the exponent. This
procedure gives  some additional information about the error of $F_{\rm
  thr}$ and $\theta$. In the
cases presented here varying $F_{\rm thr}$ by about 5\% still showed relatively good power
laws leading to changing the exponent by less than 5\% which  also serves
here to determine the error of the exponents (instead of the lower value
obtained by the statistical analysis by the regression). The values for the
exponents are $0.5\pm 0.02$ (for $\nu=0.43$ and $\nu=0.35$) and $0.49\pm 0.02$ 
(for $\nu=0.4$). 
Obviously for all densities  the exponent agrees very well with $1/2$,
suggesting a 
quadratic dependence on the drag force acting on the intruder.

\section{Conclusion}

We investigated the density dependence of the penetration behavior of a model
for collapsing ``living quicksand''. We could achieve a data collapse for
small forces when plotting the penetration depth depending on the applied
force divided by the density. For higher forces or penetration depth inertial
effects are important and the scaling is less pronounced. During the penetration
process the intruder velocity fluctuates around a constant value except for
the very initial acceleration and the final deceleration. This constant
velocity shows a power law with exponent $\theta=1/2$ as function of the
driving force minus a threshold force. This means that the drag force on the
intruder shows a quadratic velocity dependence. Here the drag force is the
driving force shifted by a threshold force.

%mean field value for the velocity exponent $\theta$ is $1/2$

\section{Acknowledgement}
We are deeply indebted to Ioannis Vardoulakis for his generosity in sharing
his ideas and the inspiring discussions about this subject. 
We thank CNPq, CAPES, FUNCAP and the PRONEX-CNPq/FUNCAP grant for
financial support.

\bibliographystyle{unsrt}
\bibliography{paper_QS_refs}

\end{document}